\documentclass[12pt,a4paper]{article}
\usepackage{amsmath}
\usepackage{amssymb}
\usepackage[cp866]{inputenc}
\usepackage[russian]{babel}

\begin{document}

\begin{center}
\textbf{Study of the Pion-Nucleon Coupling Constant Charge Dependence on the
Basis of the Low-Energy Data on Nucleon-Nucleon Interaction}\bigskip

\textbf{V. A. Babenko\footnote{%
E-mail: pet2@ukr.net} and N. M. Petrov}

\bigskip

\textit{Bogolyubov Institute for Theoretical Physics, National Academy of
Sciences of Ukraine, Kiev}
\end{center}

\thispagestyle{empty}

\vspace{1pt}

\noindent We study relationship between the physical quantities that
characterize pion-nucleon and nucleon-nucleon interaction on the basis of
the fact that nuclear forces in the nucleon-nucleon system at low energies
are mainly determined by the one-pion exchange mechanism. By making use of
the recommended proton-proton low-energy scattering parameters, we obtain
the following value for the charged pion-nucleon coupling constant g$_{\pi
^{\pm }}^{2}/4\pi =14.55(13)$. Calculated value of this quantity is in
excellent agreement with the experimental result g$_{\pi ^{\pm }}^{2}/4\pi
=14.52(26)$ of the Uppsala Neutron Research Group. At the same time, the
obtained value of the charged pion-nucleon coupling constant differs
markedly from the value of the neutral pion-nucleon coupling constant g$%
_{\pi ^{0}}^{2}/4\pi =13.55(13)$. Thus, our results show considerable charge
splitting of the pion-nucleon coupling constant.\bigskip

\bigskip

\textbf{1.} The pion-nucleon coupling constants are fundamental physical
characteristics of strong nuclear interaction. These quantities play an
important role in studying the nucleon-nucleon and pion-nucleon interactions
[1--5]. In view of this, much attention has been given for many decades to
their studying and refining their values. A detailed history of the
development of the situation around the pion-nucleon coupling constant can
be found in [5--7].

At the present time, there is no substantial controversy about the neutral
pion-nucleon coupling constant g$_{\pi ^{0}}^{2}/4\pi $. The value of g$%
_{\pi ^{0}}^{2}/4\pi =13.52(23)$ from [8], which is among the values
determined experimentally in recent years, is in perfect agreement with the
earlier results from [9], g$_{\pi ^{0}}^{2}/4\pi =13.55(13)$, and [10], g$%
_{\pi ^{0}}^{2}/4\pi =13.61(9)$, as well as with the averaged value of g$%
_{\pi ^{0}}^{2}/4\pi =13.6(3)$ quoted in [5, 11].

However, there is no unanimous consensus on the value of the charged
pion-nucleon coupling constant g$_{\pi ^{^{\pm }}}^{2}/4\pi $. The
well-known compilation of Dumbrajs and his coauthors [12] gives the value of
g$_{\pi ^{^{\pm }}}^{2}/4\pi =14.28(18)$ obtained in [13, 14] from data on $%
\pi ^{^{\pm }}p$ scattering. On the basis of an energy-dependent
partial-wave analysis (PWA) of data on nucleon-nucleon scattering, the
Nijmegen group found the value of g$_{\pi ^{^{\pm }}}^{2}/4\pi =13.54(5)$
[6, 15] for the charged pion-nucleon coupling constant. This result was
nearly coincident with coupling constant g$_{\pi ^{0}}^{2}/4\pi =13.55(13)$
determined for neutral pions by the same group in [9]. The values g$_{\pi
^{^{\pm }}}^{2}/4\pi \sim 13.7\div 13.8$ of the charged pion-nucleon
coupling constant, which are close to the constant g$_{\pi ^{0}}^{2}/4\pi
\sim 13.6$ for neutral pions, were recently obtained on the basis of data on 
$\pi ^{^{\pm }}p$ interaction in some other studies [16--19]. At the same
time, the Uppsala Neutron Research Group obtained much greater values for
the charged pion-nucleon coupling constant, g$_{\pi ^{^{\pm }}}^{2}/4\pi
=14.62(35)$ [20], g$_{\pi ^{^{\pm }}}^{2}/4\pi =14.52(26)$ [21], and g$_{\pi
^{^{\pm }}}^{2}/4\pi =14.74(33)$ [22], which exceed substantially the
average value of the coupling constant for neutral pions, g$_{\pi
^{0}}^{2}/4\pi =13.6(3)$ [5, 11]. Thus, it is of paramount importance to
address presently the problem of the possible charge dependence of the
pion-nucleon coupling constant --- in other words, the question of whether
the pion-nucleon coupling constants for neutral and charged pions differ
from each other.

In the present study, we examine the pion-nucleon coupling constants for
neutral and charged pions on the basis of data on low-energy nucleon-nucleon
($NN$) scattering. At the present time, semiphenomenological
one-boson-exchange potential models, which include the exchange of various
mesons, are frequently used to describe the $NN$ interaction. In this
approach, the exchange of pions, which are light, determines primarily the
long-range part of the $NN$ potential, while the exchange of rho and omega
mesons, which are heavier, determines the interaction at intermediate and
short distances, which is substantial at higher energies. At extremely low
energies, which effectively correspond to long distances, the use of the
simplest one-pion-exchange potentials is quite appropriate for describing
the $NN$ interaction. In view of this circumstance and under the assumption
that nuclear forces in the $NN$ system at low energies are due primarily to
the exchange of virtual pions, we use here the well-known Yukawa potential,
which directly follows from meson field theory [1--3], to describe the $NN$
interaction. In this case, we match the parameters of this potential (that
is, its depth $V_{0}$ and range $R$) with the low-energy parameters of $NN$
scattering in the $^{1}S_{0}$ spin-singlet state. By employing the
parameters of the Yukawa potential for proton-proton ($pp$) and
neutron-proton ($np$) interactions, we obtain here an equation that relates
the charged ($f_{\pi ^{^{\pm }}}^{2}$) and neutral ($f_{\pi ^{0}}^{2}$)
pseudovector pion-nucleon coupling constants to each other.

\textbf{2.} According to meson field theory, strong nuclear interaction
between two nucleons is due largely to the exchange of virtual pions, which
determine primarily the long-range part of the $NN$ interaction and,
accordingly, $NN$ scattering at extremely low energies. The nucleon-nucleon
potential that follows from meson field theory and which one calls the
Yukawa potential has the form [1--3]%
\begin{equation}
V_{Y}\left( r\right) =-V_{0}\frac{e^{-\mu r}}{\mu r}~,  \tag{1}
\end{equation}%
where $r$ is the distance between two nucleons and $\mu $ is related to the
pion mass $m_{\pi }$ by the equation%
\begin{equation}
\mu =\frac{m_{\pi }c}{\hbar }~.  \tag{2}
\end{equation}%
Here, $c$ is the speed of light and $\hbar $ is the reduced Planck constant.
The nuclear-force range $R$ is in inverse proportion to the pion mass and is
small:%
\begin{equation}
R\equiv \frac{1}{\mu }=\frac{\hbar }{m_{\pi }c}\thicksim 1.4~\text{fm}~. 
\tag{3}
\end{equation}%
As a matter of fact, the nuclear-force range $R$ is coincident with the
Compton wavelength of the pion.

The depth $V_{0}$ of the potential in (1) is related to the dimensionless
pseudovector pion-nucleon coupling constant $f_{\pi }$ by the simple
equation [1--3, 5, 23]%
\begin{equation}
V_{0}=m_{\pi }c^{2}f_{\pi }^{2}~.  \tag{4}
\end{equation}%
Thus, the pion mass $m_{\pi }$ and the pion-nucleon coupling constant $%
f_{\pi }$ are basic characteristics of the pion-nucleon interaction, which
play a significant role in studying the nucleon-nucleon and pion-nucleus
interactions [1--5]. The pseudovector coupling constants for neutral ($%
f_{\pi ^{0}}^{2}$) and charged ($f_{\pi ^{\pm }}^{2}$) pions are related to
the pseudoscalar coupling constants g$_{\pi ^{0}}^{2}$ and g$_{\pi ^{\pm
}}^{2}$ by the equations [5, 11]%
\begin{equation}
\frac{\text{g}_{\pi ^{0}}^{2}}{4\pi }=\left( \frac{2M_{p}}{m_{\pi ^{\pm }}}%
\right) ^{2}f_{\pi ^{0}}^{2}~,  \tag{5}
\end{equation}%
\begin{equation}
\frac{\text{g}_{\pi ^{\pm }}^{2}}{4\pi }=\left( \frac{M_{p}+M_{n}}{m_{\pi
^{\pm }}}\right) ^{2}f_{\pi ^{\pm }}^{2}~,  \tag{6}
\end{equation}%
\ where $M_{p}$ and $M_{n}$ are, respectively, the proton and neutron masses
and $m_{\pi ^{\pm }}$ is the charged-pion mass.

We now present those experimental values of the nucleon and meson masses and
the $\hbar c$ value that we will use in the ensuing calculations. We have%
\begin{equation}
M_{p}=938.272046~\text{MeV}/c^{2}~,~~M_{n}=939.565379~\text{MeV}/c^{2}~, 
\tag{7}
\end{equation}%
\begin{equation}
m_{\pi ^{0}}=134.9766~\text{MeV}/c^{2}~,~~m_{\pi ^{\pm }}=139.57018~\text{MeV%
}/c^{2}~,  \tag{8}
\end{equation}%
\begin{equation}
\hbar c=197.3269718~\text{MeV}\cdot \text{fm}~,  \tag{9}
\end{equation}%
where $m_{\pi ^{0}}$ is the neutral-pion mass. All of the values in Eqs.
(7)--(9) were taken from the compilation of the Particle Data Group [24].

Two charged protons interact via the exchange of a neutral pion. According
to Eqs. (2) and (4), the parameters $\mu _{pp}$ and $V_{0}^{pp}$ of the
Yukawa potential (1) are then determined by the neutral-pion mass $m_{\pi
^{0}}$ and the coupling constant $f_{\pi ^{0}}$. But in the case of
neutron-proton interaction, both neutral and charged pions are exchanged. In
the latter case, one should determine the parameters $\mu _{np}$ and $%
V_{0}^{np}$ in the potential (1) by employing [25] the averaged values of
the pion mass, 
\begin{equation}
\overline{m}_{\pi }=\frac{1}{3}\left( m_{\pi ^{0}}+2m_{\pi ^{\pm }}\right) 
\tag{10}
\end{equation}%
and of the pion-nucleon coupling constant,%
\begin{equation}
\overline{f_{\pi }^{2}}=\frac{1}{3}\left( f_{\pi ^{0}}^{2}+2f_{\pi ^{\pm
}}^{2}\right) ~.  \tag{11}
\end{equation}

In order to determine the parameters of the potential in (1), we will use
data on the interaction of two nucleons at low energies in the $^{1}S_{0}$
spin-singlet state. Further, we evaluate the proton-proton parameters $\mu
_{pp}$ and $V_{0}^{pp}$ on the basis of the scattering length $a_{pp}$ and
the effective range $r_{pp}$. In doing this, corrections associated with
electromagnetic interaction should be removed from the experimental values
of the nuclear-Coulomb low-energy parameters of $pp$ scattering. After the
removal of these corrections, the values of the purely nuclear scattering
length $a_{pp}$ and effective range $r_{pp}$ for proton-proton scattering
become [4]%
\begin{equation}
a_{pp}=-17.3(4)\,\text{fm~},  \tag{12}
\end{equation}%
\begin{equation}
r_{pp}=2.85(4)\,\text{fm~}.  \tag{13}
\end{equation}

By employing the variable-phase approach [26] and the values of the $pp$%
-scattering parameters in (12) and (13), we obtain the following results for
the parameters of the Yukawa potential (1) in the case of proton-proton
scattering:%
\begin{equation}
\mu _{pp}=0.8393\,\text{fm}^{-1}\text{~},  \tag{14}
\end{equation}%
\begin{equation}
V_{0}^{pp}=44.8295\,\text{MeV~}.  \tag{15}
\end{equation}%
In accordance with Eqs. (2), (4), (14), and (15), the mass and the
pion-nucleon coupling constant for the neutral pion in the case of the
Yukawa potential for proton-proton interaction proved to be%
\begin{equation}
m_{\pi ^{0}}^{Y}=165.6108\,\text{MeV}/c^{2}\text{~},  \tag{16}
\end{equation}%
\begin{equation}
\left( f_{\pi ^{0}}^{Y}\right) ^{2}=0.2707\text{~}.  \tag{17}
\end{equation}%
They are much greater than the experimental values%
\begin{equation}
m_{\pi ^{0}}=134.9766\,\text{MeV}/c^{2}\text{ [24]~},  \tag{18}
\end{equation}%
\begin{equation}
f_{\pi ^{0}}^{2}=0.0749(7)\text{ [9]~}.  \tag{19}
\end{equation}

Thus, we have%
\begin{equation}
m_{\pi ^{0}}^{Y}=C_{m_{\pi }}m_{\pi ^{0}}\text{~},  \tag{20}
\end{equation}%
\begin{equation}
\left( f_{\pi ^{0}}^{Y}\right) ^{2}=C_{f_{\pi }}f_{\pi ^{0}}^{2}\text{~}, 
\tag{21}
\end{equation}%
where $C_{m_{\pi }}$ and $C_{f_{\pi }}$ are constants, which, in general,
depend on the form of nucleon-nucleon interaction. From Eqs. (16)--(19), it
follows that, for the Yukawa potential, these constants are%
\begin{equation}
C_{m_{\pi }}=1.2270\text{~},  \tag{22}
\end{equation}%
\begin{equation}
C_{f_{\pi }}=3.6142\text{~}.  \tag{23}
\end{equation}

It is natural to assume that relations similar to Eqs. (20) and (21) hold
for the charged-pion mass and charged pion-nucleon coupling constants and,
hence, for the averaged pion mass and the averaged pion-nucleon coupling
constant, i.e. $C_{m_{\pi ^{0}}}=C_{m_{\pi ^{\pm }}}=C_{\overline{m}_{\pi }}$%
, $C_{f_{\pi ^{0}}}=C_{f_{\pi ^{\pm }}}=C_{\overline{f}_{\pi }}$. In view of
this, it can readily be shown that the neutron-proton parameters $\mu _{np}$
and $V_{0}^{np}$ of the potential in (1) are related to the analogous
proton-proton interaction parameters $\mu _{pp}$ and $V_{0}^{pp}$ by the
equations%
\begin{equation}
\mu _{np}=\frac{\overline{m}_{\pi }}{m_{\pi ^{0}}}\mu _{pp}\text{~}, 
\tag{24}
\end{equation}%
\begin{equation}
V_{0}^{np}=\frac{\overline{m}_{\pi }}{m_{\pi ^{0}}}\frac{\overline{f_{\pi
}^{2}}}{f_{\pi ^{0}}^{2}}V_{0}^{pp}\text{~}.  \tag{25}
\end{equation}

In accordance with Eqs. (8) and (10), the ratio of the average pion mass $%
\overline{m}_{\pi }$ to the neutral pion mass $m_{\pi ^{0}}$ is%
\begin{equation}
\frac{\overline{m}_{\pi }}{m_{\pi ^{0}}}=1.0227\text{~}.  \tag{26}
\end{equation}%
By using Eqs. (14), (24), and (26) and the experimental value of the singlet 
$np$ scattering length for the $^{1}S_{0}$ state [27, 28],%
\begin{equation}
a_{np}=-23.71(2)~\text{fm~},  \tag{27}
\end{equation}%
we obtain the following values for the parameters $\mu _{np}$ and $%
V_{0}^{np} $ in the case of $np$ interaction in the form of the Yukawa
potential:%
\begin{equation}
\mu _{np}=0.8584\,\text{fm}^{-1}\text{~},  \tag{28}
\end{equation}%
\begin{equation}
V_{0}^{np}=48.0742\,\text{MeV~}.  \tag{29}
\end{equation}%
As before, we use the variable-phase approach [26] in our calculations.

The effective range $r_{np}$ found for $np$ scattering by using the Yukawa
potential with the parameter values in (28) and (29),%
\begin{equation}
r_{np}=2.70(4)\,\text{fm~},  \tag{30}
\end{equation}%
is in perfect agreement with the experimental value [27, 28]%
\begin{equation}
r_{np}=2.70(9)\,\text{fm~}.  \tag{31}
\end{equation}%
The values of the singlet $np$ scattering length in (27) and effective range
in (30) are in very good agreement with the values of $a_{np}=-23.7154(80)~$%
fm and $r_{np}=2.706(67)\,$fm that we obtained in [29, 30] by using the
experimental values of the $np$-scattering cross section and the
experimental values of the deuteron characteristics. Thus, the use of the
experimental values of the neutral- and charged-pion masses in (8) leads to
a consistent description of experimental data on proton-proton and
neutron-proton scattering in the region of low energies.

From Eqs. (11) and (25), one can obtain an important equation that relates
the pseudovector charged and neutral pion-nucleon coupling constants.
Specifically, we have%
\begin{equation}
f_{\pi ^{\pm }}^{2}=\frac{1}{2}\left( 3\frac{V_{0}^{np}}{V_{0}^{pp}}\frac{%
\mu _{pp}}{\mu _{np}}-1\right) f_{\pi ^{0}}^{2}\text{~}.  \tag{32}
\end{equation}%
Employing Eq. (32) and taking into account the neutral pion-nucleon coupling
constant in (19), along with the parameters of proton-proton scattering in
(14) and (15) and the parameters of neutron-proton scattering in (28) and
(29), we obtain the following value for the pseudovector charged
pion-nucleon coupling constant $f_{\pi ^{\pm }}^{2}$%
\begin{equation}
f_{\pi ^{\pm }}^{2}=0.0804(7)\text{~}.  \tag{33}
\end{equation}

Employing Eqs. (5) and (6) and taking into account the pseudovector coupling
constants for the neutral pion in (19) and for the charged pions in (33),
along with the proton and neutron masses in (7) and the charged-pion mass in
(8), we obtain the following values for the pseudoscalar pion-nucleon
coupling constants:%
\begin{equation}
\frac{\text{g}_{\pi ^{0}}^{2}}{4\pi }=13.55(13)~,  \tag{34}
\end{equation}%
\begin{equation}
\frac{\text{g}_{\pi ^{\pm }}^{2}}{4\pi }=14.55(13)~.  \tag{35}
\end{equation}%
The value in (35) that we found for the pseudoscalar charged pion-nucleon
coupling constant is in perfect agreement with the experimental coupling
constant%
\begin{equation}
\frac{\text{g}_{\pi ^{\pm }}^{2}}{4\pi }=14.52(26)~,  \tag{36}
\end{equation}%
determined by the Uppsala Neutron Research Group [21].

Thus, the value in (35) obtained in the present study for the charged
pion-nucleon coupling constant differs sizably from the value of the neutral
pion-nucleon coupling constant in (34). In relative units, this difference
is about 7\%, which is indicative of a substantial charge dependence of the
pion-nucleon coupling constant.

\textbf{3.} Considering that nuclear forces in nucleon-nucleon system at low
energies are due primarily to the exchange of virtual pions, we have studied
relations between quantities that characterize the pion-nucleon and
nucleon-nucleon interactions. We have derived an equation that relates
pseudovector charged and neutral pion-nucleon coupling constants to the
depths of the potentials that describe the $np$ and $pp$ interactions. By
employing the values of $a_{pp}=-17.3(4)~$fm and $r_{pp}=2.85(4)~$fm
currently recommended for the purely nuclear $pp$ scattering length and
effective range, respectively, along with the experimental value of $%
a_{np}=-23.71(2)~$fm for the singlet $np$ scattering length and the value of
g$_{\pi ^{0}}^{2}/4\pi =13.55(13)$ for the pseudoscalar neutral pion-nucleon
coupling constant [9], we have obtained the value of g$_{\pi ^{\pm
}}^{2}/4\pi =14.55(13)$ for the charged pion-nucleon coupling constant. This
value is in very good agreement with the experimental values of g$_{\pi
^{\pm }}^{2}/4\pi =14.52(26)$ [21] and g$_{\pi ^{\pm }}^{2}/4\pi =14.74(33)$
[22], and with the value of g$_{\pi ^{\pm }}^{2}/4\pi =14.28(18)$ found
previously in [13, 14]. At the same time, the value that we obtained is at
odds with the value of g$_{\pi ^{\pm }}^{2}/4\pi =13.54(5)$ [6, 15] found by
using data on $NN$ scattering and the value of g$_{\pi ^{\pm }}^{2}/4\pi
=13.76(1)$ [16, 17] determined from data on $\pi N$ scattering. It is worth
noting, however, that values of the charged pion-nucleon coupling constant
that are smaller than g$_{\pi ^{\pm }}^{2}/4\pi =14.55(13)$ lead to the
weakening of the neutron-proton potential and to the absolute value of the
singlet $np$ scattering length underestimated in magnitude with respect to
its experimental value of $\left\vert a_{np}\right\vert =23.71(2)~$fm
[27--32].

The results obtained in the present study are indicative of a substantial
charge dependence of the pion-nucleon coupling constant. A simultaneous
analysis of low-energy pion-nucleon and nucleon-nucleon parameters leads to
the conclusion that the violation of the charge independence of nuclear
forces is due primarily to the difference in mass and in pion-nucleon
coupling constants between the neutral and charged pions. Indications of
effects stemming from the mass difference between the neutral and charged
pions and leading to a violation of the charge independence of nuclear
forces can be found in [1, 23]. Our results show that a difference of 3.4\%
in mass and a difference of 7\% in pion-nucleon coupling constant between
the neutral and charged pions lead to the difference of $pp$ and $np$
scattering lengths that is equal to $\Delta a_{\text{CIB}}\equiv
a_{pp}-a_{np}=6.41~$fm, which is about 30\% in relative units. In that case,
the difference of the effective ranges for proton-proton and neutron-proton
scattering is $\Delta r_{\text{CIB}}\equiv r_{pp}-r_{np}=0.15~$fm.

\begin{center}
REFERENCES\vspace{1pt}
\end{center}

\begin{enumerate}
\item L. Hulth\'{e}n and M. Sugawara, in \textit{Structure of Atomic Nuclei}%
, Vol. 39 of \textit{Handbuch der Physik}, Ed. by S. Fl\"{u}gge (Springer,
Berlin, G\"{o}ttingen, Heidelberg, 1957), p. 7.

\item A. Bohr and B. R. Mottelson, \textit{Nuclear Structure, }Vol. 1: 
\textit{Single-Particle Motion} (Benjamin, New York, 1969).

\item T. Ericson and W. Weise, \textit{Pions and Nuclei}, International
Series of Monographs on Physics, Vol. 74 (Oxford Univ. Press, Oxford, 1988).

\item G. A. Miller, B. M. K. Nefkens, and I. \v{S}laus, Phys. Rept. \textbf{%
194}, 1\ (1990).

\item R. Machleidt, and I. Slaus, J. Phys. G \textbf{27}, R69\ (2001).

\item J. J. de Swart, M. C. M. Rentmeester, and R. G. E. Timmermans,
arXiv:9802084 [nucl-th].

\item Proceedings of a Workshop on Critical Issues in the Determination of
the Pion-Nucleon Coupling Constant, 7-8 June 1999, Uppsala, Sweden. Phys.
Scr. \textbf{T87}, 5--77\ (2000).

\item V. Limkaisang, K. Harada, J. Nagata, \textit{et al.}, Prog. Theor.
Phys. \textbf{105}, 233\ (2001).

\item J. R. Bergervoet, P. C. van Campen, R. A. M. Klomp, \textit{et al.},
Phys. Rev. C \textbf{41}, 1435\ (1990).

\item R. A. Arndt, I. I. Strakovsky, and R. L. Workman, Phys. Rev. C \textbf{%
52}, 2246\ (1995).

\item R. Machleidt, and M. K. Banerjee, Few-Body Syst. \textbf{28}, 139\
(2000).

\item O. Dumbrajs, R. Koch, H. Pilkuhn, \textit{et al.}, Nucl. Phys. B%
\textbf{\ 216}, 277\ (1983).

\item D. V. Bugg, A. A. Carter, and J. R. Carter, Phys. Lett. B \textbf{44},
278\ (1973).

\item R. Koch, and E. Pietarinen, Nucl. Phys. A \textbf{336, }331 (1980).

\item V. Stoks, R. Timmermans, and J. J. de Swart, Phys. Rev. C \textbf{47},
512\ (1993).

\item R. A. Arndt, W. J. Briscoe, I. I. Strakovsky, \textit{et al.}, Phys.
Rev. C \textbf{69}, 035213\ (2004).

\item R. A. Arndt, W. J. Briscoe, I. I. Strakovsky, R. L. Workman, Phys.
Rev. C \textbf{74}, 045205\ (2006).

\item D. V. Bugg, Eur. Phys. J. С \textbf{33, }505 (2004).

\item V. Baru, C. Hanhart, M. Hoferichter, \textit{et al.}, Nucl. Phys. A 
\textbf{872, }69 (2011).

\item T. E. O. Ericson, B. Loiseau, J. Nilsson, \textit{et al.}, Phys. Rev.
Lett. \textbf{75}, 1046\ (1995).

\item J. Rahm, J. Blomgren, H. Cond\'{e}, \textit{et al.}, Phys. Rev. C 
\textbf{57}, 1077\ (1998).

\item J. Rahm, J. Blomgren, H. Cond\'{e}, \textit{et al.}, Phys. Rev. C 
\textbf{63}, 044001\ (2001).

\item L. A. Sliv, Izv. Akad. Nauk SSSR, Ser. Fiz. \textbf{38}, 2 (1974).

\item J. Beringer \textit{et al. }(Particle Data Group), Phys. Rev. D 
\textbf{86}, 010001\ (2012).

\item T. E. O. Ericson, and M. Rosa-Clot, Nucl. Phys. A \textbf{405, }497
(1983).

\item V. V. Babikov, \textit{Variable Phase Approach in Quantum Mechanics}
(Nauka, Moscow, 1976) [in Russian].

\item A. G. Sitenko and V. K. Tartakovskii, \textit{Lectures on the Theory
of the Nucleus} (Atomizdat, Moscow, 1972; Pergamon, Oxford, 1971).

\item W. O. Lock and D. F. Measday, \textit{Intermediate Energy Nuclear
Physics} (Methuen, London, 1970; Atomizdat, Moscow, 1972).

\item V. A. Babenko and N. M. Petrov, Phys. At. Nucl. \textbf{70}, 669
(2007); V. A. Babenko and N. M. Petrov, arXiv:0704.1024 [nucl-th].

\item V. A. Babenko and N. M. Petrov, Phys. At. Nucl. \textbf{73}, 1499
(2010).

\item R. W. Hackenburg, Phys. Rev. C\textbf{\ 73}, 044002\ (2006).

\item V. A. Babenko and N. M. Petrov, Phys. At. Nucl. \textbf{72}, 573
(2009).
\end{enumerate}

\end{document}